\documentclass[10pt,preprint]{article}
\usepackage{graphicx}
\usepackage{amsmath}
\usepackage{color}

\title{Backtracking of colloids: a multiparticle collision dynamics simulation study}
\author{M.~Belushkin $^{1}$, R.~G.~Winkler $^{2}$, G.~Foffi $^{1}$\\\\
\normalsize{$^{1}$ Institute of Theoretical Physics,}\\
\normalsize{Ecole Polytechnique F\'{e}d\'{e}rale de Lausanne (EPFL), CH-1015 Lausanne}
\\
\normalsize{$^{2}$ Theoretical Soft-Matter and Biophysics,}\\
\normalsize{Institute for Advanced Simulation, Forschungszentrum J{\"u}lich, D-52425 J{\"u}lich}
}
\date{}
\begin{document}

\maketitle

\begin{abstract}
The role of sound in the dynamics of mesoscale systems is typically neglected, since frequently the associated time scales are much smaller than all the other time scales of interest.
However, for sufficiently small objects embedded in a solvent with a sufficiently small sound velocity, sound can play a crucial role. In particular, behavior resembling viscoelasticity
has been theoretically predicted for non-viscoelastic fluids. This effect is due to the interference of the propagation of sound waves caused by the solute particle's motion and
hydrodynamic vortex formation. We demonstrate this effect, known as backtracking, in computer simulations employing the method of multiparticle collision dynamics. We systematically
study the influence of sound on the dynamics of the solute particle, and find that it disappears in the long-time limit. Thus, we confirm that sonic effects at the single-particle
level can be neglected at sufficiently long times.
\end{abstract}

\section{Introduction}
The progress in single molecule tracking techniques provides a route to study the dynamical aspects of individual particles, such as colloids, polymers, vesicles or even cells.
Specifically, insight is achieved in the influence of the environment on the particle behavior. This is of particular importance for dilute suspensions, where hydrodynamic effects
determine the transport properties of the solute particles. Here, the solvent characteristics, e.g., compressibility, are of paramount importance.
\par
Investigations of the dynamics of small objects suspended in a medium and driven by thermal fluctuations, i.e. undergoing Brownian motion, have a long fruitful history~\cite{Hanggi:2005}.
In the classical picture of Brownian motion, the dynamics of a solute particle experiencing viscous drag and random collisions with the molecules of the solvent is described by the
Langevin equation,
\begin{equation}
M \ddot x (t) = -\gamma \dot x(t) + \Gamma(t)\, .
\label{eq:langevin}
\end{equation}
Here $M$ is the particle mass, $x$ its coordinate, $\gamma$ is the solvent's viscous friction coefficient and $\Gamma(t)$ is a stochastic force which models collisions of the solute
particle with the molecules of the solvent. The stochastic force is {\it assumed} to be stationary, Markovian, and Gaussian with zero mean (white noise), i.e.
\begin{eqnarray}
\langle \Gamma(t) \rangle = 0,\\
\langle \Gamma(t) \Gamma(t^{\prime}) \rangle = 2 k_{B}T \gamma \delta(t-t^{\prime})\, .
\end{eqnarray}
\begin{table*}[t!]
\begin{tabular}{|c|c|c|c|c|c|c|}
\hline
Setup& Box size & Collision angle& Solvent density&  Collision time &  Colloid radius  &  Colloid mass\\ 
 & L [a] &$\alpha$ [deg.]  &  $\rho_{f}$ [$m/a^{3}$]  & $\delta t$ [$\sqrt{ma^{2}/k_{B}T}$] &  $R$ [$a$] & $M$ [$m$] \\ \hline
1 & 32 & 130 & 10 & 0.01, 0.02, 0.05, 0.1 &  2.0 & 335.1 \\ \hline
2 & 32 & 90  &  5   & 0.005, 0.01, 0.05 & 2.0 & 125 \\ \hline
\end{tabular}
\caption{\label{table:params}Parameters of the different setups for which simulations have been performed. This choice of parameters allows to illustrate the smooth transition
from normal behavior to backtracking.}
\end{table*}
\begin{table}[h!]
\begin{tabular}{|c|c|c|c|}
\hline
Setup&  Collision time  & Viscosity $\eta$ & Time scale \\ 
 & $\delta t$ [$\sqrt{ma^{2}/k_{B}T}$] & $[k_{B}T m / a^{4}]$ & ratio $\tau_{f}/\tau_{s}$ \\ \hline
1 & 0.01 & 82.2 & 0.3 \\
   & 0.02 & 41.2 & 0.6 \\
   & 0.05 & 16.7 & 1.5 \\
   & 0.1  & 8.7 & 3.0
\\ \hline
2 & 0.005 & 44.5& 0.3 \\
   & 0.01 & 22.3 & 0.6 \\
   & 0.05 & 4.6 & 2.8
\\ \hline
\end{tabular}
\caption{\label{table:scales}Viscosities and ratio of the viscous to the sonic time scales corresponding to the different collision times of the two setups.}
\end{table}

The Langevin equation~(\ref{eq:langevin}) is a stochastic version of Newton\rq{}s equations of motion with an effective force which models random uncorrelated collisions between the
solute particle and the solvent molecules.
\par
It is well known that this simple picture of Brownian motion is largely incomplete for objects in dilute solution, in particular for short times.
As early as in the 1960\rq{}s, Alder and Wainwright~\cite{Alder:1967,Alder:1970} observed by means of computer simulations strong deviations of the behavior of a Brownian particle
from behavior predicted by the simple Langevin equation~(\ref{eq:langevin}). These deviations arise from hydrodynamic correlations. Furthermore, Zwanzig and Bixon~\cite{Zwanzig:1975}
showed that in order to satisfy the equipartition theorem, it is necessary to account for the finite compressibility of the medium.
Since then, correlated momentum transport both by shear and by longitudinal wave propagation is known to play an important role in fluid dynamics.~\cite{Clercx:1992}.
Assuming fully-developed hydrodynamics, Stokes derived for a sphere in an incompressible fluid $\gamma=6 \pi \eta R$ for stick boundary conditions at the solute-solvent interface,
or $\gamma=4 \pi \eta R$ for slip boundary conditions in the long-time limit, based on a time-dependent response function, which relates the friction coefficient with the size of
the colloid~\cite{Stokes:1851}. Here, $R$ is the radius of the sphere and $\eta$ is the shear viscosity of the solvent.
\par
However, the Stokes friction coefficient describes the friction experienced by the colloidal particle only at long observation times. At short times, viscous and compressibility
effects strongly manifest themselves. This implies that at different stages of its motion, the solute particle experiences friction of different magnitude. It turns out possible to calculate
the resulting friction coefficient in Fourier space: the corresponding frequency-dependent expressions taking into account viscous vortex formation were first provided by Stokes~\cite{Stokes:1851,Zwanzig:1970},
and since then refined to include compressibility, an arbitrary degree of slip~\cite{Metiu:1977,Guz:1993,Espanol:1995_2,Felderhof:2005_1} and several geometries~\cite{Erbas:2010}.
\par
While viscous effects have received broad attention in the last years, sonic effects in viscous fluids are usually neglected. This can be easily understood in terms of the relevant time
scales~\cite{Dhont_book}:
\begin{itemize}
\item the sonic time scale corresponding to the time it takes a sound wave to propagate one colloid radius, $\tau_{s}=R/v_{s}$, with $v_{s}$ the velocity of sound in the solvent; for a micrometre-sized colloid in water $\tau_{s} \approx 1$ ns;
\item the inertial time scale corresponding to the characteristic time of ballistic motion of the colloid, $\tau_{B}=M/\gamma$; for the same colloid in water $\tau_{B} \approx 0.1$ $\mu$s;
\item the viscous time scale corresponding to the time of momentum diffusion by one colloidal radius, $\tau_{f}=R^{2}/\nu$, with $\nu$ the kinematic viscosity of the solvent, $\nu=\eta/\rho_{f}$, where $\rho_{f}$ is the mass density of the solvent; in the same system as above $\tau_{f} \approx 1$ $\mu$s.
\end{itemize}
Thus, in conditions typical in a soft-matter experiment the sonic time scale is many orders of magnitude smaller than the other time scales of interest, and it is therefore assumed that
sonic effects can be neglected. 
\par
However, recent experimental investigations have highlighted the role of sound in the formation of hydrodynamic correlations in many-body systems~\cite{Espanol:1995_1,Espanol:1995_2,Bartlett:2002}.
Previously, it was assumed that cross-correlations between pairs of colloidal particles would build up on the viscous time scale $\tau_{f}$, and only recently it has been shown
that instead they build up on the sonic time scale $\tau_{s}$.~\cite{Bartlett:2002}. Computer simulations revealed that the origin of this phenomenon lies in multiple scattering of sound
waves~\cite{Bakker:2002}, implying that the role of sound effectively extends to times much longer than the sonic time. Such phenomena highlight the importance of a solid theoretical
understanding of the interplay between the sonic and the viscous effects.
\par
\begin{figure}[t!]
\begin{center}
\includegraphics[width=\linewidth]{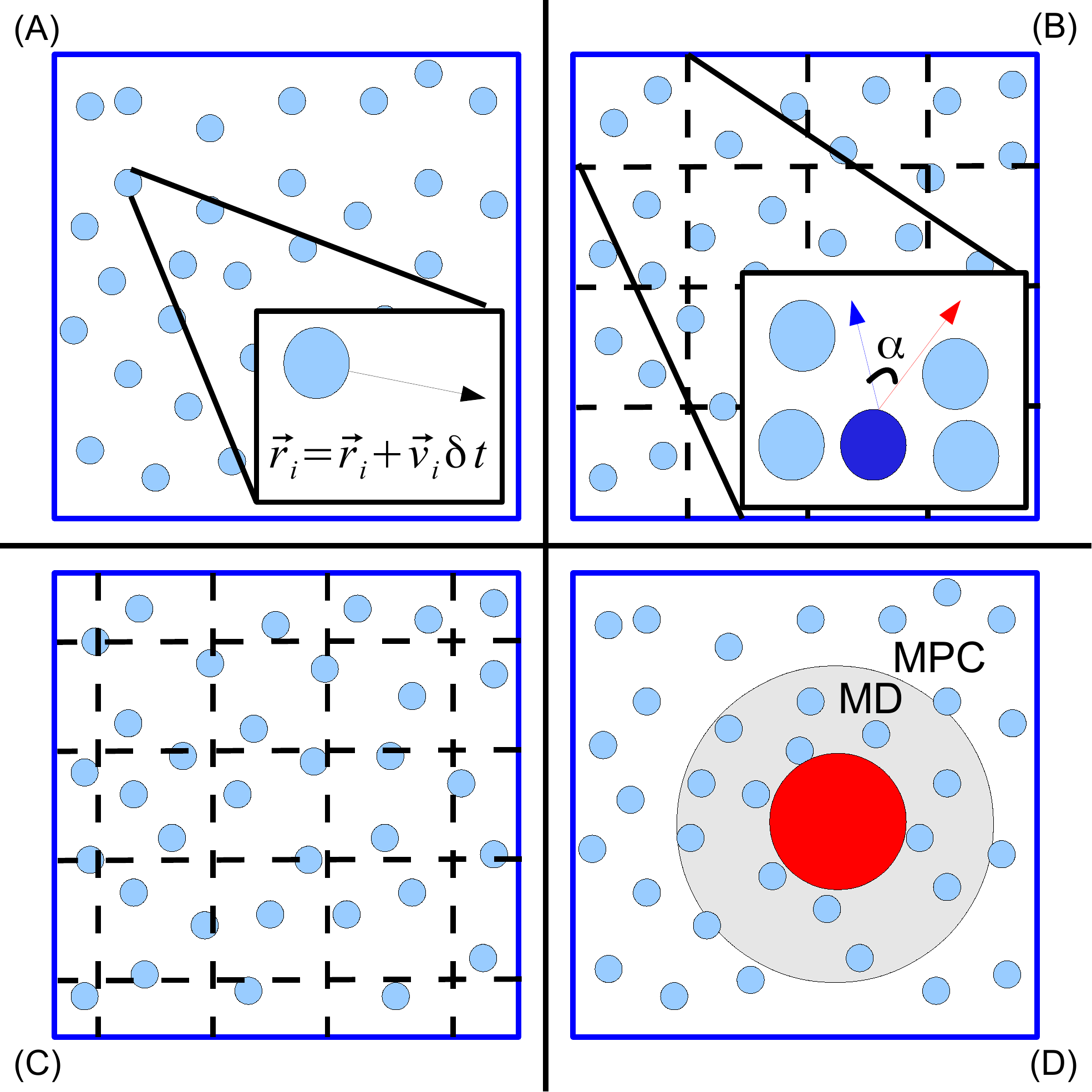}
\caption{\label{fig:mpc_steps}Schematic representation of MPC. (A) Each solvent particle undergoes streaming, whereby its coordinates are updated according to its velocity and the collision time. (B) Particles are sorted into collision cells. Within each cell, the velocity of each particle is rotated with respect to the cell\rq{}s centre-of-mass velocity by a fixed angle around a randomly oriented axis. (C) The structure of the collision grid is shifted to guarantee Galilean invariance. (D) All particles which may interact with the solute object do not undergo streaming. Instead, their equations of motion are integrated with a much finer molecular dynamics time step.}
\end{center}
\end{figure}

Even more recent theoretical investigations have revealed other ways in which sonic and viscous effects can interfere~\cite{Felderhof:2005}, demonstrating that sound propagation can give
rise to non-trivial effects not only at a finite concentration of solute particles, but also at the single-particle level. More specifically, it was shown that under conditions where the
compressibility of the fluid is sufficiently large and its bulk viscosity sufficiently small, the generation of sound waves during the motion of a single colloidal particle may lead to
a hydrodynamic reaction force which can strongly influence its future motion. This effect, known as backtracking~\cite{Felderhof:2005}, is manifested in the velocity autocorrelation function
of the colloid. Instead of a monotonic decay, it exhibits a kink at relatively small solvent compressibility, or even crosses zero at high solvent compressibility. Thus, it is predicted that
the colloid should change its initial direction of movement at times corresponding to the viscous and the sonic times, which should be of the same order of magnitude. The backtracking effect
is reminiscent of viscoelasticity, but it has a very different, dynamical origin, which can be mathematically described as follows.
\begin{figure*}[t!]
\begin{tabular}{cc}
\includegraphics[width=.5\linewidth]{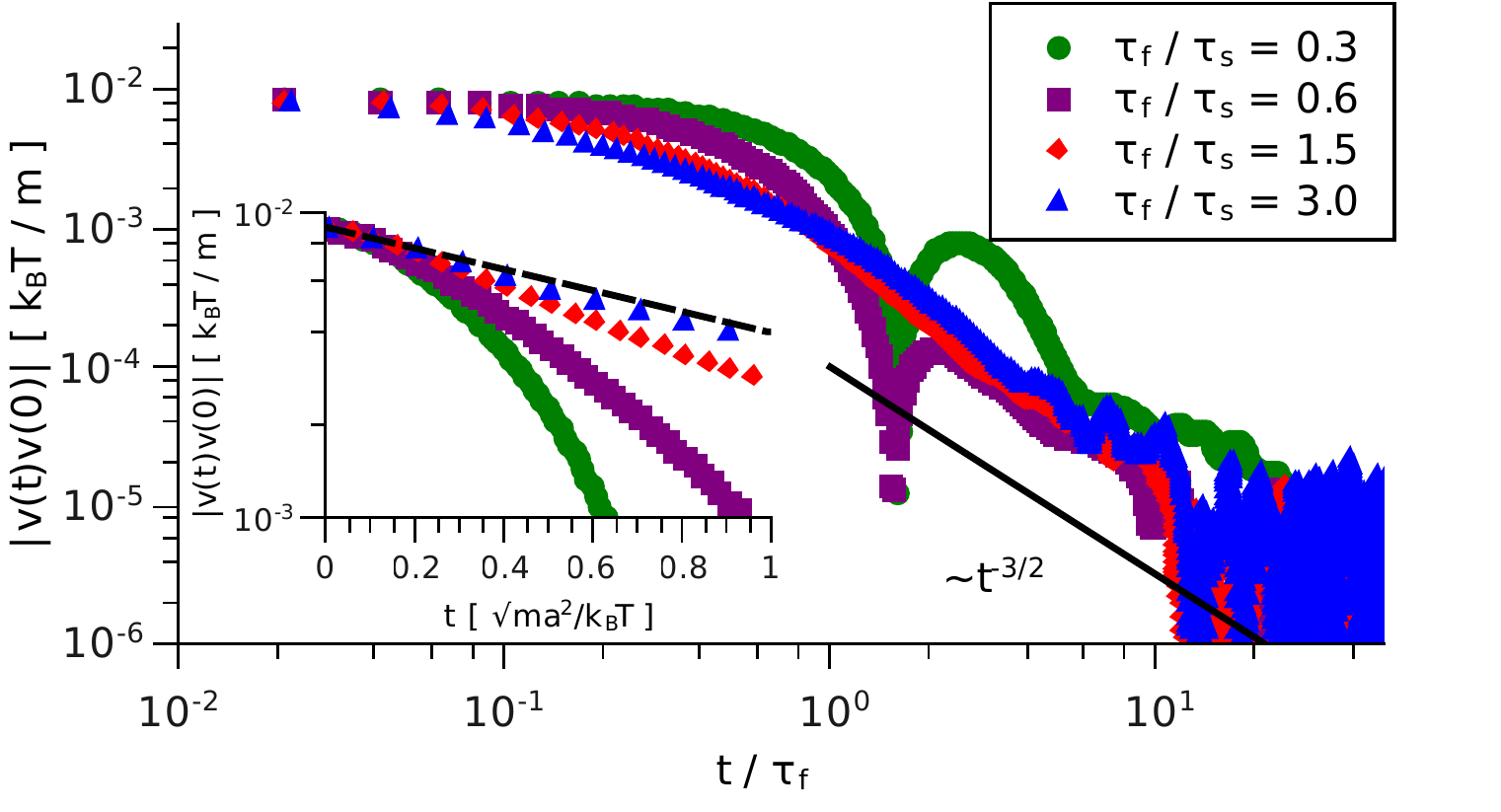}&
\includegraphics[width=.5\linewidth]{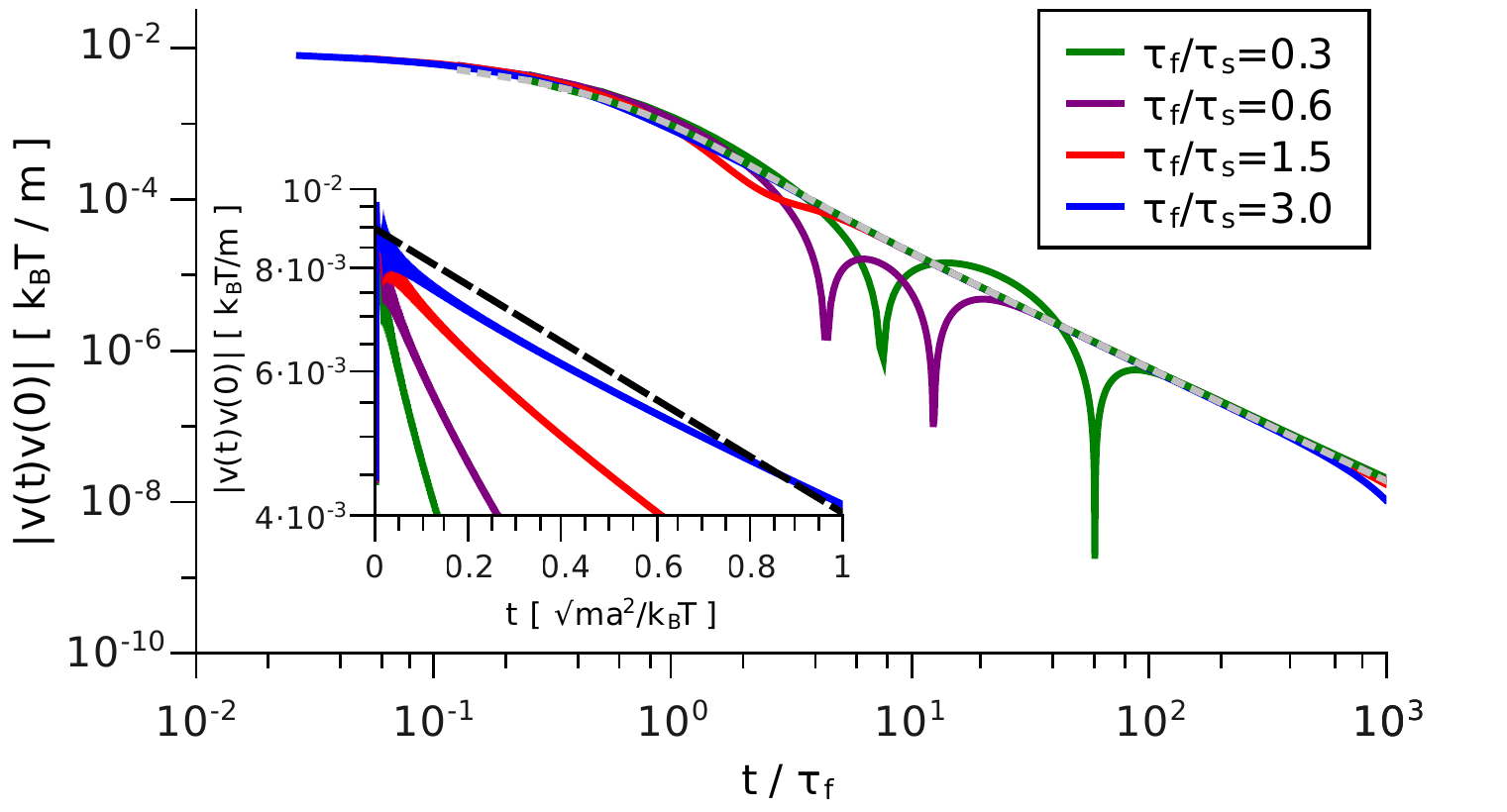}
\end{tabular}
\caption{\label{fig:rho10} Left: computer simulations for Setup 1. The solid line is a guide for the eye showing the $t^{-3/2}$ slope of the long-time tail of the velocity autocorrelation function
which indicates that hydrodynamic correlations have fully developed. The dashed line in the inset shows the Enskog exponential decay. Right: theoretical calculation for Setup 1.
Solid lines correspond to calculations including viscous effects and solvent compressibility. Dotted lines show the calculations in the incompressible limit. The dashed line in the
inset shows the Enskog exponential decay.}
\end{figure*}

The decay of the velocity autocorrelation function $C_{vv}(t)$ of a solute particle is given by
\begin{equation}
C_{vv}(t) = k_{B}T \Phi(t)
\end{equation}
with the response function
\begin{equation}
\Phi(t) = \frac{1}{2\pi} \int\limits_{-\infty}^{+\infty} {\cal Y}(\omega) {\mathrm e}^{-i\omega t} d\omega\, ,
\label{eq:response_function}
\end{equation}
where ${\cal Y}(\omega)$ is the admittance tensor,
\begin{equation}
{\cal Y}(\omega) = \frac{1}{\gamma(\omega) - i\omega M}\, ,
\end{equation}
and the frequency-dependent friction coefficient $\gamma(\omega)$ can be calculated using the analytical formulae of Refs.~\cite{Metiu:1977,Erbas:2010}. Essentially, it was
demonstrated~\cite{Felderhof:2005} that, provided that ${\cal Y}(\omega)$ exhibits certain analytical properties in the complex plane, velocity reversal of the solute particle will
occur. It was shown that this velocity reversal is due to the elasticity of the fluid corresponding to the finite velocity of sound.
\par
In this contribution, we address the question to what extent sonic effects contribute to the dynamics of colloids on the single-particle level. To this end, we study backtracking in
computer simulations which capture thermal fluctuations, hydrodynamic vorticity and sound propagation. We investigate the influence of sound on long-time observables, namely
the diffusion coefficient of a colloidal particle suspended in a solvent. The results are timely, since time scales in the range of the typical hydrodynamic time scales in soft-matter systems -
typically of the order of microseconds - are becoming more and more accessible experimentally.

\section{Simulation Method}
\begin{figure*}[t!]
\begin{tabular}{cc}
\includegraphics[width=.5\linewidth]{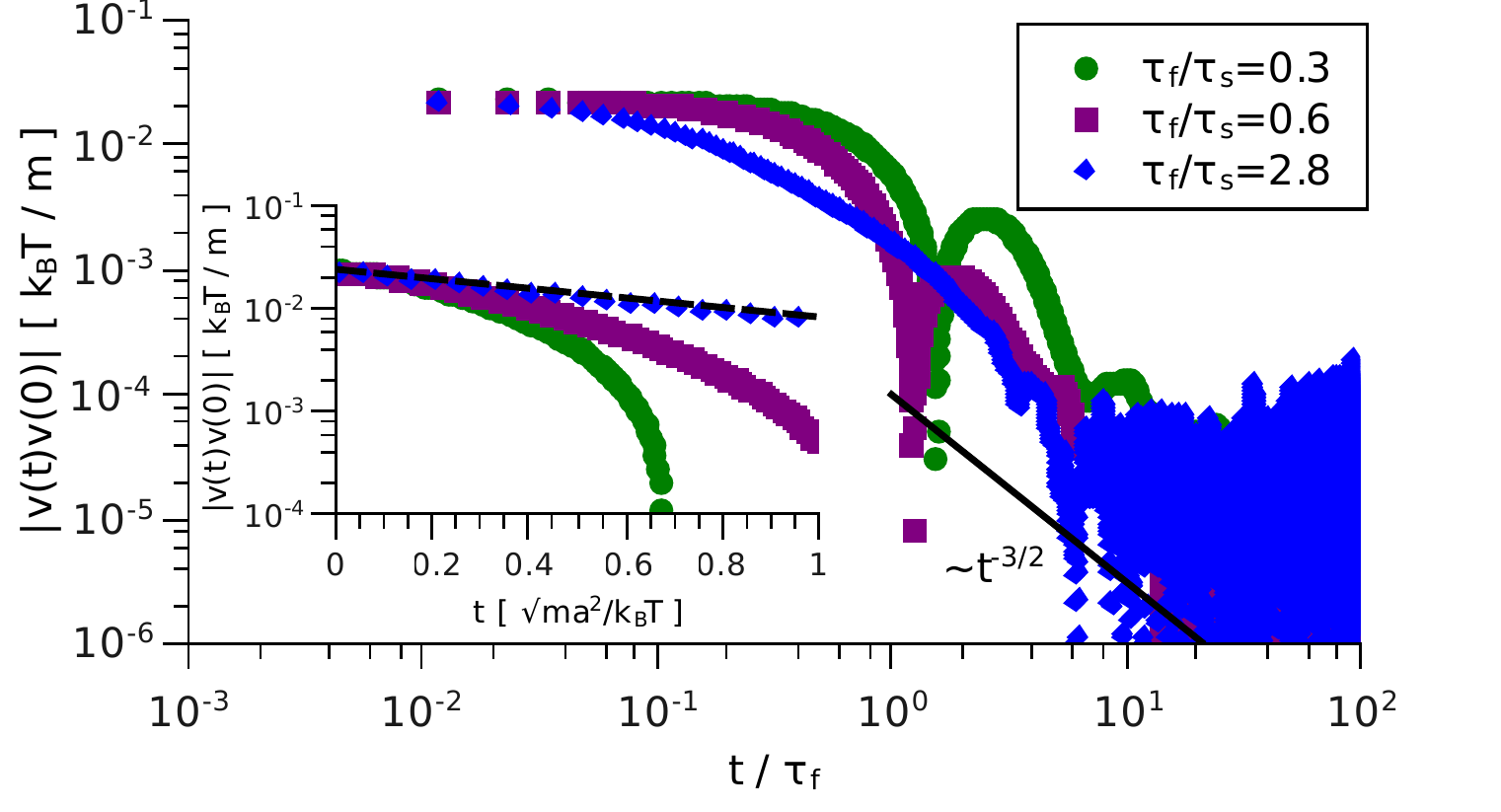}&
\includegraphics[width=.5\linewidth]{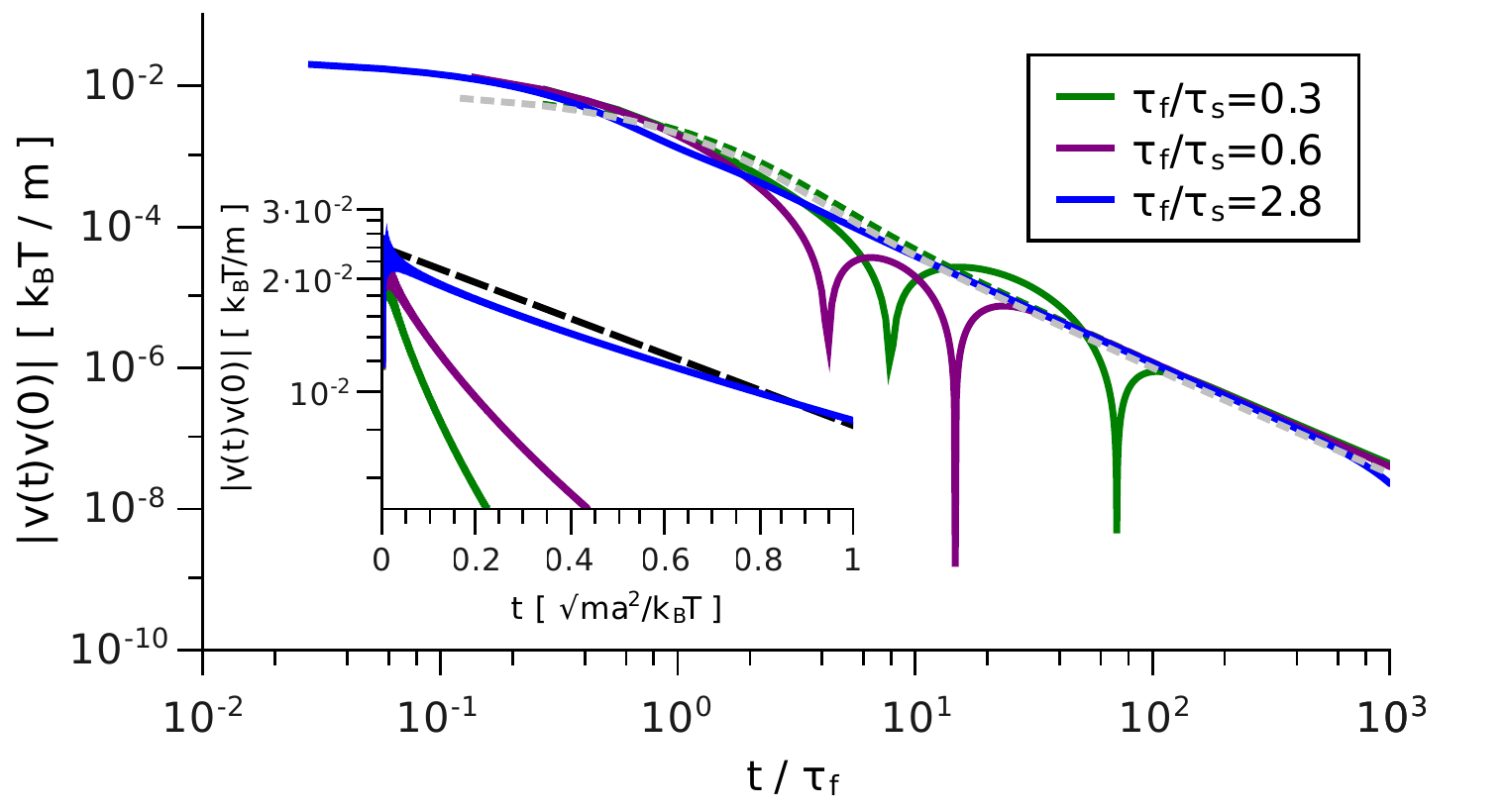}
\end{tabular}
\caption{\label{fig:rho5} Left: computer simulations for Setup 2. The solid line is a guide for the eye showing the $t^{-3/2}$ slope of the long-time tail of the velocity autocorrelation
function which indicates that hydrodynamic correlations have fully developed. The dashed line in the inset shows the Enskog exponential decay.
Right: theoretical calculation for Setup 2. Solid lines correspond to calculations including viscous effects and solvent compressibility. Dotted lines show the calculations
in the incompressible limit. The dashed line in the inset shows the Enskog exponential decay.}
\end{figure*}

We employ the method of multiparticle collision dynamics (MPC) to investigate backtracking and its influence on the long-time observables in a simple system of a spherical colloid embedded
in a viscous solvent. MPC is known to correctly capture both thermal fluctuations and hydrodynamics on coarse-grained scales~\cite{Padding:2006,Kapral:2008,Gompper:2009}, has an 
H-theorem~\cite{Malevanets:1999} and satisfies the fluctuation theorem~\cite{Belushkin:2011}. MPC is ideally suited for observation of backtracking since it satisfies the necessary
criteria~\cite{Felderhof:2005}: the bulk viscosity is zero and the sound velocity $v_{s}$ is small, since the MPC fluid behaves as an ideal gas. Therefore, the simulations present ideal
conditions for the observation of the interference of sonic and viscous effects, since the sonic, viscous and inertial time scales are very similar.
\par
To model the solvent in the simulations, a cubic box of lateral size $L$ with periodic boundary conditions is filled with $N$ pointlike solvent particles of mass $m$ with continuous
coordinates $\vec{r}_{i}$ and velocities $\vec{v}_{i}$, $i \in [1;N]$. The average kinetic energy per solvent particle determines the thermal energy of the solvent.
At each step of the simulation, the solvent undergoes streaming and collisions. During streaming, the coordinates of each solvent particle are updated according to its velocity
$\vec{v}_{i}$ and the collision time $\delta t$: $\vec{r}_{i} \rightarrow \vec{r}_{i} + \vec{v}_{i}\delta t$ (\ref{fig:mpc_steps}~(A)). Then, all particles are sorted into collision
cells of lateral size $a$ according to their positions within the simulation box (\ref{fig:mpc_steps}~(B)). The collision cells form a cubic collision grid spanning the whole simulation
box; in order to guarantee Galilean invariance, the structure of this collision grid is shifted at each simulation step~\cite{Gompper:2009,Ihle:2001} (\ref{fig:mpc_steps}~(C)).
Within a collision cell, the velocities of the particles are rotated with respect to the cell's centre-of-mass velocity by a fixed rotation angle $\alpha$ around a randomly oriented axis.
Energy and momentum are thus conserved locally. Due to the existence of an H-theorem~\cite{Malevanets:1999}, the velocity distribution of the solvent particles relaxes to a Maxwell-Boltzmann
distribution in just a few simulation steps, regardless of the initial distribution of velocities.
\par
We couple the solute colloidal particle to the solvent employing the standard hybrid molecular dynamics approach~\cite{Malevanets:2000}:
solvent particles interact with the solute through a shifted Lennard-Jones potential with a cutoff radius corresponding to the colloidal radius $R$. This form of coupling implies slip
boundary conditions at the solute-solvent interface. Stick boundary conditions can also be studied with MPC~\cite{Luijten:2010,Goetze:2010}.
\par
We choose the solute-solvent interaction potential to be
\begin{equation} U(r) =
\begin{cases}
 4 \left( \left(\displaystyle \frac{\displaystyle R}{\displaystyle r}\right)^{48} -
\left(\displaystyle \frac{\displaystyle R}{\displaystyle r}\right)^{24}\right) + 1 & ,r\leq 2^{1/24}R,\\
0 & ,r > 2^{1/24}R.
\end{cases}
\end{equation}
All solvent particles which may interact with the colloid within a MPC step do not undergo streaming. Instead, their equations of motion are integrated with the smaller molecular dynamics (MD)
time step, taking into account the solvent-solute interaction potential (\ref{fig:mpc_steps}~(D)). Afterwards, they are sorted into collision cells and undergo collision with the rest of the MPC
solvent particles.
\par
We choose the following normalization: solvent particle mass $m=1$, thermal energy $k_{B}T=1$ and collision cell size $a=1$.
Thus, mass is measured in units of $m$, energy in units of $k_{B}T$, length in units of $a$ and time in units of $t_{0}=\sqrt{ma^{2}/k_{B}T}$. We perform simulations for several setups.
The corresponding setup-specific parameters are listed in \ref{table:params}. For the observation of backtracking, it is important that the sonic time scale is not too small, i.e. the ratio of
viscous to sonic time scales is not large. In \ref{table:scales} we list the shear viscosities corresponding to each of the time steps of the two setups; the viscosities are calculated using
the analytical expressions of Refs.~\cite{Kapral:2008,Gompper:2009,Ihle:2005,Kikuchi:2003,Noguchi:2008,Winkler:2009}. The corresponding ratios of the viscous time scale
$\tau_{f}=R^{2}\rho_{f}/\eta$ to the sonic time scale $\tau_{s}=R/v_{s}$ are also shown. Since our fluid obeys ideal-gas equations of state, the velocity of sound - and, consequently, the sonic
time - is fixed, i.e., $v_{s}=\sqrt{5/3}$~\cite{Gompper:2009,Huang:2010}. Therefore, it is the viscous time scale which changes between the
different simulations due to the difference in viscosities which, in our case, depend on the collision time $\delta t$, collision angle $\alpha$, and the solvent density $\rho_{f}$.
\section{Results}
The colloid velocity autocorrelation functions (VACFs), averaged over several hundred simulation runs each, are shown in \ref{fig:rho10} and \ref{fig:rho5} for Setup 1 and Setup 2, respectively.
The corresponding theoretical curves are calculated in terms of the inverse Fourier transform of the admittance tensor~\cite{Felderhof:2005} (\ref{eq:response_function}) with the frequency-dependent
friction coefficient calculated following Ref.~\cite{Erbas:2010}. A key ingredient of this calculation is the hydrodynamic radius of the colloidal particle $R_{H}$, such that in the
zero-frequency limit the Stokes relation is satisfied, $\gamma(\omega=0) = 4\pi \eta R_{H}$. Equivalently, the theoretical calculation can be performed if simply the long-time limit of
the friction coefficient is known. This is the approach we adopt: we assume the validity of the Stokes-Einstein relation at long times, $D = k_{B}T / \gamma$, and measure the corresponding
friction coefficient $\gamma$ in simulations.
\par
The measured data are in good qualitative agreement with the theory and, clearly, several regimes can be distinguished.
\begin{figure}[t!]
\begin{tabular}{cc}
\includegraphics[width=.5\linewidth]{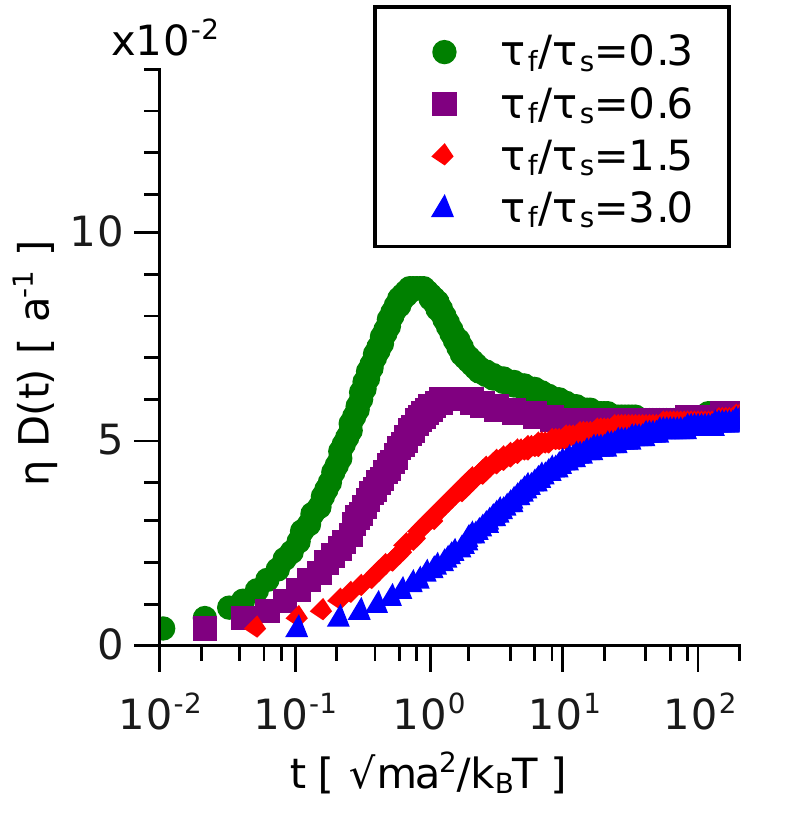} &
\includegraphics[width=.5\linewidth]{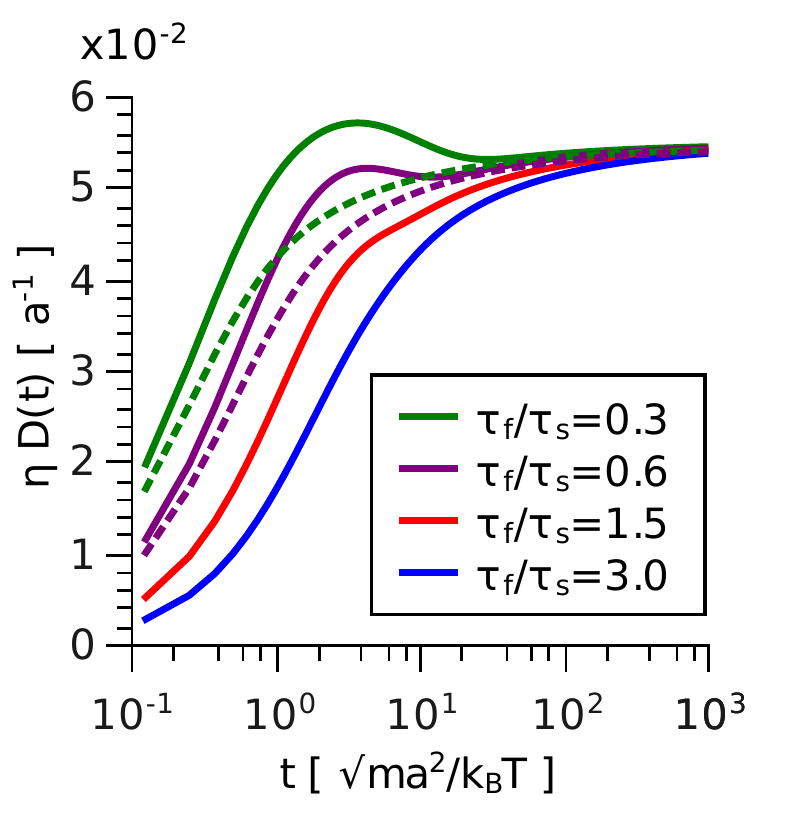}
\end{tabular}
\caption{\label{fig:rho10_diffusion} The product of the solvent viscosity $\eta$ and the time-dependent diffusion coefficient $D(t)$ calculated using \ref{eq:doft} in simulations (left) and
theory (right) for Setup 1. Dotted lines show the results of the calculations in the incompressible limit. In all cases, the long-time limit of the diffusion coefficient is unaffected by the
effects of sound.}
\end{figure}
\par
First of all, we observe that at sufficiently long times, all the velocity autocorrelation functions exhibit the same long-time tail which depends only on the properties of the
solvent~\cite{Hauge:1973,Hinch:1975,Paul:1981,Ernst:1970,Padding:2006},
\begin{equation}
\lim_{t\rightarrow \infty} \langle v(t) v(0) \rangle \approx \frac{k_{B}T}{12 \rho_{f} (\pi \nu t)^{3/2}}\, .
\end{equation}
In this limit, the simulations are in excellent agreement with the theory. We again stress that the long-time tail is a direct consequence of the Navier-Stokes equation which yields the
time-dependent (or, equivalently, the frequency-dependent) friction coefficient. The Langevin equation itself predicts a simple exponential decay of the VACF.
\par
\begin{figure}[t!]
\begin{tabular}{cc}
\includegraphics[width=.5\linewidth]{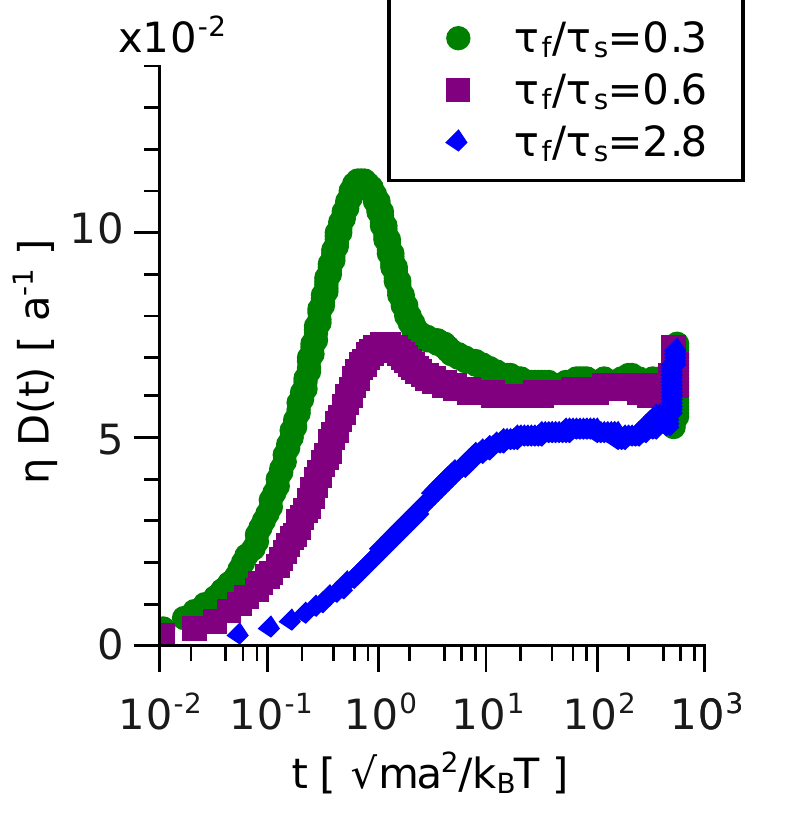} &
\includegraphics[width=.5\linewidth]{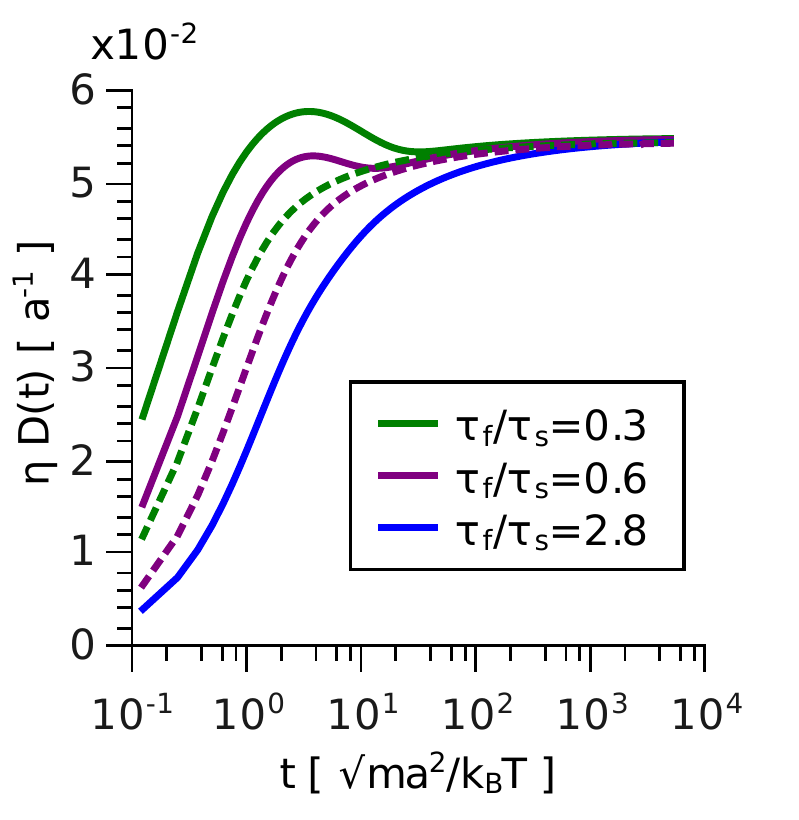}
\end{tabular}
\caption{\label{fig:rho5_diffusion} The product of the solvent viscosity $\eta$ and the time-dependent diffusion coefficient $D(t)$ calculated using \ref{eq:doft} in simulations (left) and
theory (right) for Setup 2. Dotted lines show the results of the calculations in the incompressible limit. In all cases, the long-time limit of the diffusion coefficient is unaffected by the
effects of sound.}
\end{figure}

At very short times, however, there is an apparent discrepancy between the simulations and the theory, as indicated in the insets of \ref{fig:rho10} and \ref{fig:rho5}. Whereas simulations
exhibit an exponential decay of the velocity autocorrelation function, the theory predicts a decay which is faster than exponential. This difference has a straightforward explanation. At
very short times, the solvent in the simulations can not be treated as a continuous medium, and the Navier-Stokes equation can not be applied. Instead, one has to assume random collisions between
solvent particles and the colloid. Here, the Langevin equation applies, and the velocity autocorrelation function indeed decays exponentially, with the decay constant given by the ratio of
the Enskog friction $\gamma_{E}$ and the colloid mass $M$~\cite{Padding:2006}, with
\begin{equation}
\gamma_{E}=\frac{8}{3}\left( \frac{2 \pi k_{B}T M m}{M+m}\right)^{1/2} \frac{\rho_{f}}{m} R^{2}\, .
\label{eq:enskog_gamma}
\end{equation}
This behavior is, in principle, a consequence of the coarse-graining approach of MPC - the relatively small number of solvent particles and the discrete dynamics imply that few collisions
occur in a single simulation step, and, consequently, a continuum description is not applicable.
We remark that the Enskog friction coefficient has a different physical origin than the Stokes viscous friction coefficient, and that the two are not necessarily related.
\par
The theory, on the other hand, assumes that a hydrodynamic description is valid down to arbitrarily short times. The Enskog regime is completely absent from the theoretical treatment.
Consequently, the theoretical curves differ only when compressibility is accounted for, and only at the sonic time scale (see \ref{fig:rho10}, right).
\par
We further observe that for characteristic sonic times $\tau_{s}$ smaller than viscous times $\tau_{f}$, the velocity autocorrelation functions decay monotonically and
do not exhibit any peculiar properties.  However, when $\tau_{s}$ becomes larger than $\tau_{f}$, anticorrelations clearly develop in the velocity autocorrelation function,
signaling backtracking. The monotonic decay of the VACF is superseded by crossing zero and an asymptotic approach of the hydrodynamic $t^{-3/2}$ long-time tail. Here, the
simulation results are in good qualitative agreement with the predictions of the theory. We note that if, in the theoretical calculations, the effects of solvent compressibility
are not taken into account, i.e. if the calculations are performed with an infinitely high velocity of sound, no anticorrelations develop (dotted lines of \ref{fig:rho10} and \ref{fig:rho5}).
Therefore, these anticorrelations are indeed a direct consequence of the interference of sonic and viscous effects.
\par
\begin{figure}[t!]
\begin{center}
\includegraphics[width=\linewidth]{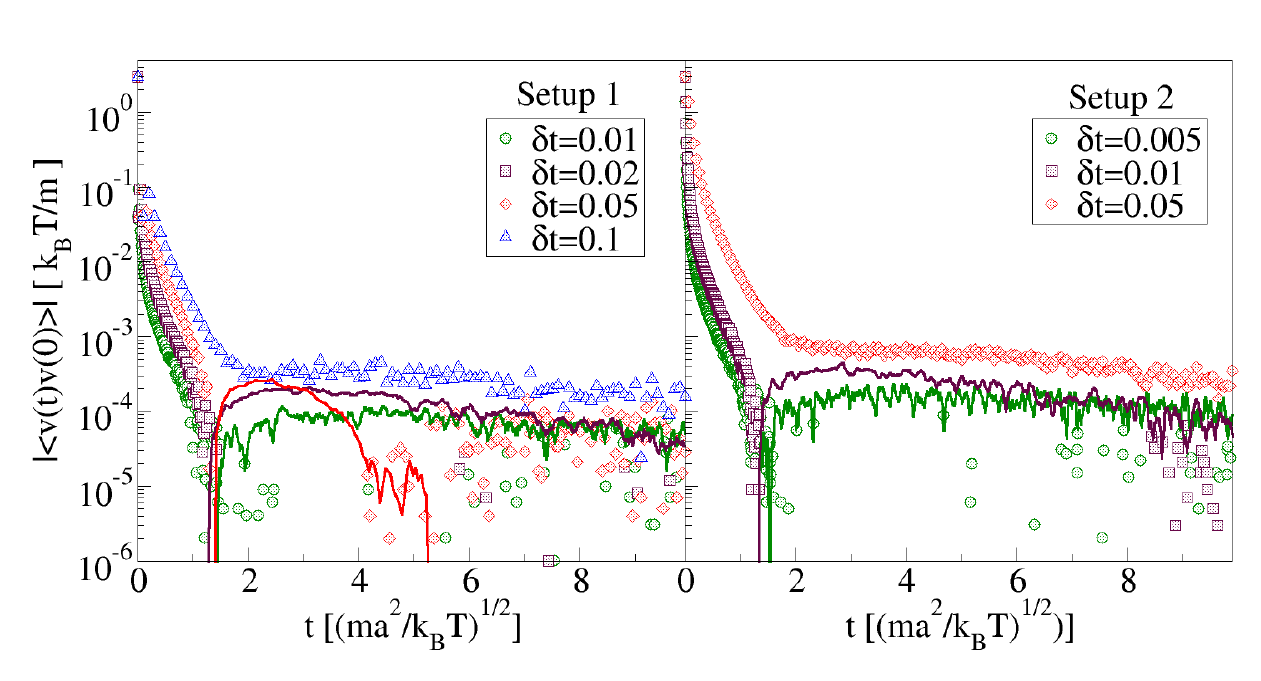}
\end{center}
\caption{\label{fig:solvent}Velocity autocorrelation functions for solvent particles for $\alpha=130^{\mathrm o}$ and $\rho=10$ (left) and $\alpha=90^{\mathrm o}$ and $\rho=5$ (right).
The points correspond to positive values of the VACF and the solid lines correspond to negative values of the VACF.\label{fig:solvent_vacf}}
\end{figure}

To study the influence of backtracking on long-time transport properties, we calculate the diffusion coefficient as the integral of the velocity autocorrelation function, i.e.,
\begin{equation}
D(t) = \int\limits_{0}^{t} \langle v(\tau)v(0)\rangle d\tau\, .
\label{eq:doft}
\end{equation}
Assuming the validity of the Stokes-Einstein relation,
\begin{equation}
D = \frac{k_{B}T}{\gamma} = \frac{k_{B}T}{4 \pi \eta R}\,
\end{equation}
where $D\equiv D(\infty)$, the product $D\eta$ depends only on the properties of the colloid, namely its size, and does not depend on properties of the solvent, namely its shear viscosity,
and serves as a good indicator of the relevance of sonic effects at long times.
\par
As shown in \ref{fig:rho10_diffusion} and \ref{fig:rho5_diffusion}, we observe that in the presence and in the absence of backtracking, $\eta D(t)$ tends to the same value in the asymptotic limit.
Clearly, at the sonic time scale, there are significant deviations between the cases with and without backtracking. However, the long-time limit is always the same.
As before, the simulation results are in good qualitative agreement with theory. We note that if the solvent compressibility is eliminated from the theoretical calculations, the long-time
limit of $\eta D(t)$ remains the same (dotted lines of \ref{fig:rho10_diffusion} and \ref{fig:rho5_diffusion}).
\par
Hence, we have confirmed that even when the effects of sound are very pronounced and dramatically influence the short-time regime of the solvent particle\rq{}s motion,
the net effect remains zero, i.e. the role of sound can be neglected when studying long-time observables.
\par
The main requirement to observe backtracking is that the sonic time scale should be equal to or slightly larger than the viscous time scale. This is clearly not the case in
typical conditions of a micron-sized particle suspended in a liquid like water since, as we demonstrated above, in this case the sonic time scale is of the order of nanoseconds, whereas
the viscous time scale is of the order of microseconds. However, the sonic time scale is proportional to the size of the particle, whereas the viscous time scale is
proportional to the square of the size of the particle, and already for a particle a few nanometres in size, the sonic time scale and the viscous time scale have the same order of
magnitude - namely, several picoseconds. Clearly, the viscous time scale can further be reduced by employing a more viscous fluid, and conditions under which backtracking occurs
can be reached experimentally.
\par
To complete the analysis, we study the velocity autocorrelation function of the solvent itself~\cite{Ripoll:2005}. For small time steps, it is known to exhibit anticorrelations, and it may be
tempting to assume that anticorrelations in the velocity of the solute are a direct consequence of anticorrelations in the velocities of the solvent particles. However, we find that
anticorrelations within the solvent and anticorrelations of the colloidal particle do not necessarily arise together, as can be seen by comparing the data for
$\tau_{f}/\tau_{s}=1.5$ in \ref{fig:rho10} and \ref{fig:solvent_vacf} (left). 

\section{Conclusions}
We have confirmed that interference of sound and viscous effects leads to a nonmonotonic decay of the velocity autocorrelation function of a single colloidal particle suspended in a viscous solvent.
Furthermore, we have verified that under certain conditions backtracking can occur, i.e. the velocity autocorrelation function becomes negative. This behavior is reminiscent of viscoelastic
behavior - however, it has a very different, dynamical origin which lies in the interference of the formation of sound waves and hydrodynamic vorticity. Finally, we have confirmed that
despite the strong influence of backtracking on the velocity autocorrelation function, its long-time integral, the diffusion coefficient, is unaffected. That is to say that at long times the
effects of sound are completely integrated out. An important consequence of this result is the confirmation that sonic effects at the single-particle level can be neglected at long observation times.

\section{Acknowledgements}
MB and GF acknowledge financial support by the Swiss National Science Foundation (grant no. PP0022\_119006). GF acknowledges Prof. Gene Stanley for several scientific and non-scientific discussions.

\end{document}